\documentclass[12pt]{article}
\usepackage[mathscr]{eucal}
\usepackage{epsfig,amsfonts}
\usepackage[fleqn]{amsmath}
\usepackage{amsthm,amssymb}
\usepackage{graphicx}
\usepackage{hhline}
\usepackage{cite}
\usepackage{mathrsfs}
\usepackage{amsmath,amssymb}
\usepackage{hyperref}
\usepackage{verbatim}
\usepackage{stmaryrd}
\usepackage{amsfonts}
\usepackage{amsmath,amssymb}

\usepackage{amsfonts}
\usepackage{amscd}
\usepackage{bbm}

\makeatletter \@addtoreset{equation}{section} \makeatother

\newcommand{\alg}[1]{\mathfrak{#1}}

\newcommand{\comm}[2]{[#1,#2]}
\newcommand{\acomm}[2]{\{#1,#2\}}
\newcommand{\nn}{\nonumber}

\begin{document}

\begin{titlepage}
\begin{flushright}
DMUS-MP-12-02\\
NORDITA-2012-11\\
\end{flushright}

\begin{center}
\begin{LARGE}

{\bf Multi-parametric R-matrix for the $\mathfrak{sl}(2|1)$
Yangian}

\end{LARGE}
\vspace{1.3cm}

\begin{large}

{\bf  Andrei Babichenko}$^{1}$ and {\bf  Alessandro
Torrielli}$^{2}$

\end{large}

\vspace{1.cm}

${}^{1}$Department of Particle Physics, Weizmann Institute,
Rehovot 76100, Israel \\ \vspace{.2cm}

${}^{2}$Department of Mathematics, University of Surrey,
Guildford, Surrey, GU2 7XH, UK \\
\vspace{1.0cm}

\end{center}

\vspace{2cm}
\begin{center}
\centerline{\bf Abstract} \vspace{.8cm}
\parbox{15.5cm}
{ We study the Yangian of the $\mathfrak{sl}(2|1)$ Lie
superalgebra in a multi-parametric four-dimensional
representation. We use Drinfeld's second realization to {\rm
independently rederive} the R-matrix, {\rm and to obtain} the
antiparticle representation, the crossing and the unitarity
condition. We consistently apply the Yangian antipode and its
inverse to the individual particles involved in the scattering. We
explicitly find a scalar factor solving the crossing and unitarity
conditions{}{, and study the analytic structure of the resulting
dressed R-matrix}. The formulas we obtain bear some similarities
with those familiar from the study of integrable structures in the
AdS/CFT correspondence, although they present obvious crucial
differences. }

\end{center}

\vfill

\end{titlepage}
\newpage

\section{Introduction}

In recent years there has been an increasing interest in
integrable models based on superalgebra symmetries, both in the
continuum and in their lattice versions. Examples of this sort
include spin chains on one hand, and integrable and conformal
field theories on the other hand, most notably two-dimensional
sigma models on supergroup manifolds. The spectum of spin chains
with superalgebra symmetries turns out to be interesting, in
particular, in view of their conformal limit
{}{\cite{Maas,EFS,GM,FM,Can}}. This limit is expected to reproduce
the data obtained by elaborated methods of logarithmic conformal
field theories (for introductory reviews see e.g. \cite{G,F}).
Integrability makes it in principle possible to determine the spin
chain spectrum {}{exactly}, and therefore can allow to derive
exact predictions for the CFT spectrum and partition function in
the conformal limit.

\medskip

The interest in two-dimensional sigma models on supergroup
manifolds emerged both from string theory {}{\cite{BZV,BVW,BPR}}
and in context of disordered two-dimensional condensed matter
systems \cite{BL,GLL}. Later on, sigma models on a variety of
supergroup manifolds, and their Gross-Neveu like analogs, were
successfully investigated by integrability methods. Especially
interesting is the relation of integrable structures to the CFT
ones, when the sigma model is not only integrable but also
conformal with a non chiral conformal symmetry \cite{BZV,Bab}.

\medskip

One instance where the integrability based on superalgebras
revealed itself particularly powerful is in the investigation of
AdS/CFT correspondence for maximally supersymmetric backgrounds.
In the case of integrable backgrounds the same R-matrix appears in
their sigma model and spin chain incarnations, on the AdS and CFT
side respectively, and enables an exact comparison of the
quantities on both sides of correspondence. For a review see e.g.
\cite{Bei} and references therein. In the case of the $AdS_5\times
S^5$ background, {\rm the R-matrix scatters two excitations transforming in the fundamental representation of two copies of the centrally extended $\mathfrak{psl}(2|2)$
algebra \cite{Beisert:2005tm}. In the case of $AdS_4\times
CP(3)$, the R-matrix scatters two types of excitations (A and B) transforming under    $\mathfrak{psl}(2|2)$.} Another case of an integrable background where
alternating spin chain{}{s} seems to be relevant is $AdS_3\times
S^3 \times S^3 \times S^1$ \cite{SS,BSZ}. 
{\rm The superconformal algebras on which the spin chains are based upon in the three models mentioned above are $\mathfrak{psl}(4|4)$, $\mathfrak{osp}(6|4)$ and $D(2,1;\alpha)^2$, respectively.}
In all these cases the
relevant superalgebra representations depend on some continuum
parameters, which enter the non-relativistic dispersion relation
of the excitations in the system. The dependence of the
transfer-matrix spectrum on such additional parameters in the spin
chain case, or, equivalently, its dependence on the {}{`particle'}
mass spectrum in the sigma model case, raises the important
physical question of their interpretation in the framework of
integrability. This question has been raised earlier in {}{the}
literature \cite{Bazha}.

\medskip

In this paper we consider another example of rational R-matrix
based on the superalgebra $\mathfrak{sl}(2|1)$, taken in a four
dimensional representation (and its conjugated). These
representations may be considered as fundamental and
anti-fundamental for an $\mathfrak{osp}(2|2)$ algebra, which is
isomorphic to $\mathfrak{sl}(2|1)$. R- and S- matrices in this
representations and with this symmetry were considered earlier in
the literature \cite{Maas,BL,Gru1,Gru2,GM,SP}, see also the recent
paper \cite{FM}, but in a different setup and {\rm with a
different methodology. Our approach allows us to independently
rederive for the convenience of our purposes\footnote{\label{f1}
\rm R-matrices similar to the ones we obtain here from the Yangian
construction can most likely be derived by taking the rational
limit of the quantum affine result $R^{V_4 \, V_4'}$ in
\cite{Gru2}.} the R-matrix $R$ (see text). Most importantly, the
Yangian construction enables us to perform a thorough study of
crossing transformations and the related overall dressing phase.}
We construct these R-matrices explicitly from the requirement of
their commutation with the $\mathfrak{sl}(2|1)$ Yangian
comultiplication, and show that they satisfy {}{the Yang-Baxter
equation}. Each four dimensional representation of
$\mathfrak{sl}(2|1)$ corresponds to a point in moduli space which
depends on five parameters {\rm ($a-e$ in what follows)} related by two constraints, such that
the R-matrix depends on three parameters for each of the two
representations it intertwines. 
{\rm A further similarity transformation relates this representation to an equivalent one with a single continuum parameter, related to the eigenvalue of the $\alg{gl}(1)$ part of the even subalgebra. However, we choose to leave the parameters $a-e$ explicit in our treatment, as it is convenient for comparisons ({\it e.g.} with \cite{Gru2} and with the AdS/CFT literature) and at the same time it is relatively easy to do with the Yangian machinery.} The R-matrices we find depend on
the difference of the Yangian spectral parameters carried by each
representation, and in a non-difference form on all remaining
parameters. The conjugation rule of representation involves a non
trivial change of representation parameters. We succeed {}{in
finding} a relativistic interpretation to these conjugation
transformations as antiparticles, such that the R-matrix we find
is crossing invariant and unitary. In order to achieve this, the
R-matrix needs to be multiplied by a crossing-unitarizing scalar
factor, which we find explicitly. An interesting effect is
revealed concerning the role of the inverse of the antipode in the
crossing relation.

\medskip

Our multi-parametric R-matrix{}{\footnote{For related work, see
\cite{Bracken:1994hz,Delius:1994ab}.}} is therefore a good
candidate for describing an integrable two-dimensional sigma model
based on the $\mathfrak{sl}(2|1)$ superalgebra. One of the
possible candidates is the `supersymmetric {\it sign} Gordon'
(SSSG) model on the super manifold
$\mathfrak{osp}(3|2)/\mathfrak{osp}(2|2)$. For some recent
developments on the SSSG see \cite{HHM}. We also produce a formal
interpretation of the parameters characterizing our
representations in terms of {}{the} variables used in the context
of integrability of the AdS/CFT correspondence. The R-matrix we
find resembles very closely Beisert's R-matrix
{}{\cite{Beisert:2005tm}}, being however different. We
nevertheless believe that our findings might be instrumental in
resolving certain issues, related for instance to the possibility
of a Drinfeld's second realization of the AdS/CFT Yangian in the
distinguished basis \cite{Sp}. In this paper, we derive such a
realization for a similar four-dimensional representation
(although of a different superalgebra), and show how the
supercharges get modified at the Yangian level by the presence of
the multi-parametric deformation. This turns out to be quite
similar to how certain so-called `secret' charges, found in
\cite{Matsumoto:2007rh}, appear in the AdS/CFT context. Such
charges might therefore be related to an alternative choice of a
Dynkin diagram with respect to the one in \cite{Spill:2008tp}, and
connected to it by commutation with the secret `automorphism'
generator $\widehat{\mathfrak{B}}$.

\medskip

The plan of the paper is the following. In section \ref{section2},
we study the Yangian of $\mathfrak{sl}(2|1)$ in the distinguished
basis and in the four-dimensional representation relevant to our
interests. We utilize Drinfeld's second realization of the
Yangian, and derive the R-matrix in this representation. We also
check the Yang-Baxter equation and the unitarity condition, and
expand the R-matrix in terms of projectors onto irreducible
components of the tensor product of two four-dimensional
$\mathfrak{sl}(2|1)$ representation. In section \ref{section3}, we
derive the conjugate representation and its R-matrix, and then
derive the antiparticle representation. We also comment on the
similarities with the AdS/CFT case. In section \ref{section4}, we
derive the antiparticle R-matrix and the crossing symmetry
condition, which we explicitly solve obtaining a crossing
symmetric and unitary scalar factor.  We consistently apply the
Yangian antipode and its inverse separately on the two factors of
the tensor product, and study its effect on the
particle-antiparticle transformation. We finish with some
conclusions, {}{an appendix with formulas for the conjugate
R-matrix, and another appendix with the analysis of the poles of
the direct R-matrix (dressed with the scalar factor)}.

\section{R matrix from the Yangian}\label{section2}

Let us denote with $E_{ij}$ the matrix with all zeroes, but $1$ in
row $i$, column $j$. We will work with a so-called {\it
distinguished} Dynkin diagram, {}{{\it i.e.} with the lowest
number of fermionic nodes} (in this case, one, {\rm corresponding to generators $E_2$, $F_2$ and $H_2$ below}). The representation
we are interested in is the following:

\begin{eqnarray}
\label{algebra}
&&E_1 = E_{43}, \qquad F_1 = E_{34}, \qquad H_1 = - E_{33} + E_{44},\\
&&E_2 = - a \, E_{14} + b E_{32}, \qquad F_2 = - d \, E_{23} + \,
e E_{41}, \qquad H_2 = - c \, \mathbbmss{1} - E_{11} -
E_{44},\nonumber
\end{eqnarray}
where the parameters are constrained as

\begin{eqnarray}
\label{costri} a e = c + 1, \qquad b d = c.
\end{eqnarray}

The vector space on which this representation acts is generated by
two bosons $|a\rangle$ (indices $a=1,2$) and two fermions
$|\alpha\rangle$ (indices $|\alpha\rangle=3,4$). Notice that {\rm at the values $e=0$ (corresponding to $c = -1$) and $b=0$ (corresponding to $c=0$)} the above representation {\rm is} reducible
but indecomposable. In fact, if we {\rm choose $e
= 0$}, the state $|1\rangle$ gets annihilated by all generators,
but the state $|4\rangle$ is still sent to $|1\rangle$ by $E_2$. {\rm If instead we choose $b=0$, then the state $|2\rangle$ gets annihilated by all generators,
but the state $|3\rangle$ is still sent to $|2\rangle$ by $F_2$.}
The Cartan matrix, whose entries we denote with $a_{ij}$, is a two
by two matrix with entries equal to $2$ and $0$ respectively on
the diagonal, and $-1$'s on the anti-diagonal. The following
assignment\footnote{We will always denote with $[A,B]$ the graded
commutator $AB - (-)^{deg(A) deg(B)} \,B$, and with $\{A,B\}$ the
combination $AB + (-)^{deg(A) deg(B)} \,B$. The grading is $0$ for
the bosonic indices $1,2$ and $1$ for the fermionic indices $3,4$,
so that $deg(E_{ij}) = deg(i)+deg(j)$.}

\begin{eqnarray}
\label{10}
&&\xi^+_{i,0} = E_i, \qquad \xi^-_{i,0} = F_i, \qquad \kappa_{i,0} = H_i,\nonumber\\
&&\xi^+_{1,1} = u \, E_1, \qquad \xi^-_{1,1} = u \, F_1, \qquad \kappa_{1,1} = u \, H_1,\nonumber\\
&&\xi^+_{2,1} = b \, \Big(u + \frac{1}{2}\Big) \, E_{32} \, - a \, \Big(u - \frac{1}{2}\Big) \, E_{14}, \qquad \xi^-_{2,1} = e \, \Big(u - \frac{1}{2}\Big) \, E_{41} \, - d \, \Big(u + \frac{1}{2}\Big) \, E_{23}, \nonumber\\
&&\kappa_{2,1} = - c \, \Big(u + \frac{1}{2}\Big) \, \mathbbmss{1}
+ \Big(c - u + \frac{1}{2}\Big) \, (E_{11} + E_{44}),
\end{eqnarray}
with the rest of the generators $\xi^\pm_{i,n}$, $\kappa_{i,n}$,
$n>1$, consistently obtained by subsequent application of the
relations (\ref{def:drinf2}) below to the above generating
elements (\ref{10}), defines a representation of the Yangian in
Drinfeld's second realization \cite{Drin,Spill:2008tp}:

\begin{align}
\label{def:drinf2}
&[\kappa_{i,m},\kappa_{j,n}]=0,\quad [\kappa_{i,0},\xi^\pm_{j,m}]=\pm a_{ij} \,{ \xi^\pm_{j,m}},\nonumber\\
& \comm{{ \xi^+_{i,m}}}{\xi^-_{j,n}}=\delta_{i,j}\, \kappa_{j,m+n},\nonumber\\
&[\kappa_{i,m+1},\xi^\pm_{j,n}]-[\kappa_{i,m},\xi^\pm_{j,n+1}] = \pm \frac{1}{2} a_{ij} \{\kappa_{i,m},\xi^\pm_{j,n}\},\nonumber\\
&\comm{\xi^\pm_{i,m+1}}{\xi^\pm_{j,n}}-\comm{\xi^\pm_{i,m}}{\xi^\pm_{j,n+1}} = \pm\frac{1}{2} a_{ij} \acomm{\xi^\pm_{i,m}}{\xi^\pm_{j,n}},\nn\\
&i\neq j,\, \, \, \, n_{ij}=1+|a_{ij}|,\, \, \, \, \, Sym_{\{k\}}
[\xi^\pm_{i,k_1},[\xi^\pm_{i,k_2},\dots [\xi^\pm_{i,k_{n_{ij}}},
\xi^\pm_{j,l}]\dots]]=0.
\end{align}
One can actually go further, and prove that the all-level
representation corresponding to (\ref{10}) and which solves all
the relations (\ref{def:drinf2}) is given by

\begin{eqnarray}
\label{10f}
&&\xi^+_{1,n} = u^n \, E_1, \qquad \xi^-_{1,n} = u^n \, F_1, \qquad \kappa_{1,n} = u^n \, H_1,\\
&&\xi^+_{2,n} = b \, \Big(u + \frac{1}{2}\Big)^n \, E_{32} \, - a \, \Big(u - \frac{1}{2}\Big)^n \, E_{14}, \qquad \xi^-_{2,n} = e \, \Big(u - \frac{1}{2}\Big)^n \, E_{41} \, - d \, \Big(u + \frac{1}{2}\Big)^n \, E_{23}, \nonumber\\
&&\kappa_{2,n} = -a  e \, \Big(u - \frac{1}{2}\Big)^n \, E_{11}  -
b  d \, \Big(u + \frac{1}{2}\Big)^n \, E_{22}  -b d \, \Big(u +
\frac{1}{2}\Big)^n \, E_{33}  -a e\, \Big(u - \frac{1}{2}\Big)^n
E_{44}.\nonumber
\end{eqnarray}

The $R$-matrix related to this Yangian representation must have
very specific properties. Since it must satisfy

\begin{eqnarray}
\label{inv} \Delta^{op} ({\mathfrak{J}}) \, R = R \, \Delta
({\mathfrak{J}})
\end{eqnarray}
for any generator $\mathfrak{J}$ of the Yangian, we can obtain
strong constraints on its entries by focusing for instance  on the
Cartan subalgebra $\{\kappa_{i,0}, i=1,2\}$. The coproduct in this
subalgebra is trivial (as it is trivial on the entire level $n=0$
Lie subalgebra of the Yangian), namely

\begin{eqnarray}
\Delta(\kappa_{i,0}) = [\kappa_{i,0}]_{\mbox{rep 1}} \otimes
\mathbbmss{1} + \mathbbmss{1} \otimes [\kappa_{i,0}]_{\mbox{rep
2}} = \Delta^{op}(\kappa_{i,0}),
\end{eqnarray}
with $\otimes$ being the {\it graded} tensor product, such that
$(X \otimes Z) (Y \otimes W) = (-)^{deg(Z) deg(Y)} \, XY \otimes
ZW$ among operators and $(X \otimes Z) (v_1 \otimes v_2) =
(-)^{deg(Z) deg(v_1)} \, Xv_1 \otimes Zv_2$ when acting on states.

By looking at (\ref{algebra}), and recalling that the total number
of particles is conserved, we immediately obtain for example the
conservation of the following numbers:

\begin{itemize}
\item `Total number of bosons of type $1$' \ {\it minus} \ `Total number of bosons of type $2$'

\item `Total number of fermions of type $3$' \ {\it minus} \ `Total number of fermions of type $4$'
\end{itemize}

Notice that the $\pm c \, \mathbbmss{1}$ term in the Cartan
generators $H_i$ simply drops out of the relation (\ref{inv}). The
conservation of the above quantum numbers is enough to single out
the structure of the $R$-matrix entries, which must be as follows
(we denote by ${\mathfrak{ij}}$ the state $|i\rangle \otimes
|j\rangle $ for simplicity\footnote{We remind that the grading of
the states is then $deg(\mathfrak{1})=deg(1)=0$,
$deg(\mathfrak{2})=deg(2)=0$, $deg(\mathfrak{3})=deg(3)=1$,
$deg(\mathfrak{4})=deg(4)=1$.}, and we choose a specific overall
normalization):

\begin{eqnarray}
\label{Rmatrice}
&&R \, {\mathfrak{11}} = {\mathfrak{11}},\nn\\
&&R \, {\mathfrak{12}} = B \, {\mathfrak{12}} + C \, {\mathfrak{21}} + D \, {\mathfrak{34}} + E \, {\mathfrak{43}},\nn\\
&&R \, {\mathfrak{21}} = F \, {\mathfrak{12}} + G \, {\mathfrak{21}} + H \, {\mathfrak{34}} + I \, {\mathfrak{43}},\nn\\
&&R \, {\mathfrak{22}} = L \, {\mathfrak{22}},\nn\\
&&R \, {\mathfrak{33}} = \Gamma \, {\mathfrak{33}},\nn\\
&&R \, {\mathfrak{34}} = P \, {\mathfrak{12}} + Q \, {\mathfrak{21}} + N \, {\mathfrak{34}} + \Theta \, {\mathfrak{43}},\nn\\
&&R \, {\mathfrak{43}} = T \, {\mathfrak{12}} + U \, {\mathfrak{21}} + \Psi \, {\mathfrak{34}} + \Xi \, {\mathfrak{43}},\nn\\
&&R \, {\mathfrak{44}} = V \, {\mathfrak{44}},
\end{eqnarray}
and

\begin{eqnarray}
\label{Rmatrice1}
&&R \, {\mathfrak{13}} = \alpha_1 \, {\mathfrak{13}} + \alpha_2 \, {\mathfrak{31}},\nn\\
&&R \, {\mathfrak{14}} = \alpha_3 \, {\mathfrak{14}} + \alpha_4 \, {\mathfrak{41}},\nn\\
&&R \, {\mathfrak{23}} = \alpha_5 \, {\mathfrak{23}} + \alpha_6 \, {\mathfrak{32}},\nn\\
&&R \, {\mathfrak{24}} = \alpha_7 \, {\mathfrak{24}} + \alpha_8 \, {\mathfrak{42}},\nn\\
&&R \, {\mathfrak{31}} = \beta_1 \, {\mathfrak{13}} + \beta_2 \, {\mathfrak{31}},\nn\\
&&R \, {\mathfrak{41}} = \beta_3 \, {\mathfrak{14}} + \beta_4 \, {\mathfrak{41}},\nn\\
&&R \, {\mathfrak{32}} = \beta_5 \, {\mathfrak{23}} + \beta_6 \, {\mathfrak{32}},\nn\\
&&R \, {\mathfrak{42}} = \beta_7 \, {\mathfrak{24}} + \beta_8 \,
{\mathfrak{42}}.
\end{eqnarray}
Incidentally, these are the same non-zero entries of Beisert's
R-matrix {}{\cite{Beisert:2005tm}}.

One can relate some of these entries {}{to one another} by
imposing invariance under the generators $E_1$ and $F_1$ ({}{which
we call the `fermionic'} $\mathfrak{sl}(2)$ subalgebra). However,
the algebra being $\mathfrak{sl}(1|2)$, no $\mathfrak{sl}(2)$
subalgebra is available for the bosonic states, therefore many of
the coefficients of the $R$-matrix still remain unconstrained.

The comultiplication becomes non-trivial as soon as we move to the
level one Yangian generators. One has

\begin{eqnarray}
\label{coprid}
&&\Delta(\kappa_{2,1}) = \kappa_{2,1} \otimes \mathbbmss{1} + \mathbbmss{1} \otimes \kappa_{2,1} + H_2 \otimes H_2 + F_1 \otimes E_1 - F_3 \otimes E_3,\nonumber\\
&&\Delta(\kappa_{1,1}) = \kappa_{1,1} \otimes \mathbbmss{1} + \mathbbmss{1} \otimes \kappa_{1,1} + H_1 \otimes H_1 - 2 F_1 \otimes E_1 + F_2 \otimes E_2 + F_3 \otimes E_3,\nonumber\\
&&\Delta(\xi^+_{2,1}) = \xi^+_{2,1} \otimes \mathbbmss{1} + \mathbbmss{1} \otimes \xi^+_{2,1} + H_2 \otimes E_2 + F_1 \otimes E_3,\nonumber\\
&&\Delta(\xi^-_{2,1}) = \xi^-_{2,1} \otimes \mathbbmss{1} + \mathbbmss{1} \otimes \xi^-_{2,1} + F_2 \otimes H_2 - F_3 \otimes E_1,\nonumber\\
&&\Delta(\xi^+_{1,1}) = \xi^+_{1,1} \otimes \mathbbmss{1} + \mathbbmss{1} \otimes \xi^+_{1,1} + H_1 \otimes E_1 - F_2 \otimes E_3,\nonumber\\
&&\Delta(\xi^-_{1,1}) = \xi^-_{1,1} \otimes \mathbbmss{1} +
\mathbbmss{1} \otimes \xi^-_{1,1} + F_1 \otimes H_1 + F_3 \otimes
E_2,
\end{eqnarray}
where we denote the {}{generators associated to the} non-simple
roots as

\begin{eqnarray}
E_3 = [E_1,E_2], \qquad F_3 = [F_1,F_2].
\end{eqnarray}
We have checked that these coproducts provide a homomorphism of
the Yangian, namely, they respect the relations
(\ref{def:drinf2}).

By imposing the condition (\ref{inv}) and using formulas
(\ref{coprid}), together with the remaining level zero coproducts

\begin{eqnarray}
\Delta(\xi^\pm_{i,0}) = \xi^\pm_{i,0} \otimes \mathbbmss{1} +
\mathbbmss{1} \otimes \xi^\pm_{i,0} = \Delta^{op}(\xi^\pm_{i,0}),
\end{eqnarray}
one is able to fix the R-matrix entries uniquely up to an overall
scalar factor. If we define
\begin{eqnarray}
\delta u = u_1 - u_2,
\end{eqnarray}
then one finds {\rm (see also footnote (\ref{f1}))}

\begin{eqnarray}
&&B = \frac{(\delta u + c_1 - c_2)(1 + \delta u + c_1 - c_2)}{(- 1 + \delta u - c_2)(\delta u - c_2)}, \qquad C = \frac{b_2 (1 + c_2) d_1 e_1}{(- 1 + \delta u - c_2)(\delta u - c_2) e_2},\nonumber\\
&&D = - E = - \frac{b_2 (\delta u + c_1 - c_2) e_1}{(- 1 + \delta u - c_2)(\delta u - c_2)}, \qquad F = \frac{a_1 b_1 (1 + c_2) d_2}{a_2 (- 1 + \delta u - c_2)(\delta u - c_2)},\nonumber\\
&&G = \frac{\delta u (1 + \delta u)}{(- 1 + \delta u - c_2)(\delta u - c_2)}, \qquad H = - I = - \frac{\delta u \, b_1 (1 + c_2)}{a_2 (- 1 + \delta u - c_2)(\delta u - c_2)}, \nonumber\\
&&L = \frac{(\delta u + c_1) (1 + \delta u + c_1)}{(- 1 + \delta u - c_2)(\delta u - c_2)}, \qquad \Gamma = \frac{(1 + \delta u + c_1)}{(- 1 + \delta u - c_2)}, \qquad \Psi = \Theta, \qquad \Xi = N,\nonumber\\
&&N = \frac{\delta u (\delta u + c_1 - c_2)}{(- 1 + \delta u - c_2)(\delta u - c_2)}, \qquad \Theta = \frac{\delta u - c_2 (1 + c_1)}{(- 1 + \delta u - c_2)(\delta u - c_2)},\nonumber\\
&&P = - \frac{a_1 d_2 (\delta u + c_1 - c_2)}{(- 1 + \delta u - c_2)(\delta u - c_2)}, \qquad Q = - U = - \frac{d_1 \, \delta u \, (1 + c_2)}{(- 1 + \delta u - c_2)(\delta u - c_2) e_2},\nonumber\\
&&V = \Gamma, \qquad T=-P,
\end{eqnarray}
\begin{eqnarray}
&&\alpha_1 = \alpha_3 = \frac{\delta u + c_1 - c_2}{- 1 + \delta u - c_2}, \qquad \alpha_2 = \alpha_4 = \frac{a_2 (1 + c_1)}{a_1 (1 - \delta u + c_2)},\nonumber\\
&&\alpha_5 = \alpha_7 = \frac{\delta u (1 + \delta u + c_1)}{(- 1 + \delta u - c_2)(\delta u - c_2)}, \qquad \alpha_6 = \alpha_8 = \frac{b_1 (1 + \delta u + c_1) d_2}{(- 1 + \delta u - c_2)(\delta u - c_2)},\nonumber\\
&&\beta_1 = \beta_3 = \frac{(1 + c_1) e_2}{(1 - \delta u + c_2) e_1}, \qquad \beta_2 = \beta_4 = \frac{\delta u }{- 1 + \delta u - c_2},\\
&&\beta_5 = \beta_7 = \frac{b_2 (1 + \delta u + c_1) d_1}{(- 1 +
\delta u - c_2)(\delta u - c_2)}, \qquad \beta_6 = \beta_8 =
\frac{(1 + \delta u + c_1)(\delta u + c_1 - c_2)}{(- 1 + \delta u
- c_2)(\delta u - c_2)}\nonumber .
\end{eqnarray}

We have checked that the R-matrix satisfies the graded Yang-Baxter
equation

\begin{eqnarray}
&&R_{\mathfrak{i1 \, i2 \, j1 \, j2}} (x_1,x_2) \, R_{\mathfrak{j1 \, i3 \, m1 \, n3}} (x_1,x_3) \, R_{\mathfrak{j2 \, n3 \, m2 \, m3}} (x_2,x_3) \, (-)^{deg(\mathfrak{j2})(deg(\mathfrak{i3})+ deg(\mathfrak{n3}))} =\nonumber\\
&&R_{\mathfrak{i2 \, i3 \, j2 \, j3}} (x_2,x_3) \, R_{\mathfrak{i1
\, j3 \, n1 \, m3}} (x_1,x_3) \, R_{\mathfrak{n1 \, j2 \, m1 \,
m2}} (x_1,x_2) \, (-)^{deg(\mathfrak{j2})(deg(\mathfrak{j3})+
deg(\mathfrak{m3}))},
\end{eqnarray}
where {}{all indices run from $1$ to $4$ and repeated indices are
summed over. We} have defined

\begin{eqnarray}
R \, {\mathfrak{ij}} = R_{\mathfrak{i j m n}} (x_1, x_2) \,
{\mathfrak{mn}}
\end{eqnarray}
using the notation of (\ref{Rmatrice}) for the states, and
collectively indicating the representation parameters in
representation $i$ as

\begin{eqnarray}
x_i \equiv \{a_i,b_i,c_i,d_i,e_i,u_i \},
\end{eqnarray}
constrained by (\ref{costri}).

We have also checked that the above R-matrix satisfy the unitarity
condition

\begin{eqnarray}
(-)^{\mathfrak{cd} +\mathfrak{ab}} \, \, R_{\mathfrak{b a c d}}
(x_2, x_1) \, R_{\mathfrak{d c p q}} (x_1, x_2) \, = \,
\delta_{\mathfrak{a,p}} \, \delta_{\mathfrak{b,q}}.
\end{eqnarray}
This implies that any overall scalar factor multiplying this
R-matrix will have to satisfy unitarity on its own, namely.

\begin{eqnarray}
\label{unita} \Phi_{12} \, \Phi_{21} \, = 1.
\end{eqnarray}

Notice that the tensor Casimir of the algebra is given by:

\begin{eqnarray}
C_{12} = &&- \frac{1}{2} E_3 \otimes F_3 + \frac{1}{2} E_2 \otimes F_2 + \frac{1}{2} F_3 \otimes E_3 - \frac{1}{2} F_2 \otimes E_2 - \frac{1}{2} F_1 \otimes E_1 - \frac{1}{2} E_1 \otimes F_1 \nonumber\\
&&+ H_2 \otimes H_2  + \frac{1}{2} (H_1 \otimes H_2 + H_2 \otimes
H_1).
\end{eqnarray}
It satisfies {}{$[C_{12},\Delta(\mathfrak{J})]=0$} for any level
zero generator $\mathfrak{J}$. Since the level zero of the Yangian
has a trivial coproduct, the R-matrix can be decomposed into a
linear combination of projectors onto irreducible representations
of the tensor product of representations $1$ and $2$. The
irreducible components correspond to the eigenspaces of the
Casimir operator, and there are three such eigenspaces,
corresponding to the three distinct eigenvalues of $C_{12}$
{}{\cite{Beisert:2005tm,Janik:2006dc}}

\begin{eqnarray}
\label{CasEig} \lambda_1 = c_1 \, c_2, \qquad \lambda_2 = (1 +
c_1)(1 + c_2), \qquad \lambda_3 = \frac{1}{2}(c_1 + c_2 + {}{2} \,
c_1 \, c_2).
\end{eqnarray}
The projectors onto the three eigenspaces are given by

\begin{eqnarray}
\label{poiet} P_i \, = \, \frac{(C_{12} - \lambda_j)(C_{12} -
\lambda_k)}{(\lambda_i - \lambda_j)(\lambda_i -
\lambda_k)}\label{proj}
\end{eqnarray}
with $(i,j,k)=(1,2,3),(2,1,3),(3,1,2)$, respectively. The R-matrix
(without the overall scalar factor $\Phi_{12}$) can then be
written as\footnote{\rm See also footnote (\ref{f1}), and
\cite{PF,L}.}

\begin{eqnarray}
R =  \frac{(u_1 - u_2 + c_1)(1 + u_1 - u_2 + c_1)}{(-1 + u_1 - u_2
- c_2)(u_1 - u_2 - c_2)} \, P_1 \, + \, P_2 \, + \, \frac{1 + u_1
- u_2 + c_1}{- 1 + u_1 - u_2 - c_2} \, P_3.
\end{eqnarray}
The various coefficients in the above spectral decomposition
correspond to the diagonal action of the R-matrix on the highest
weight states in each irreducible component. Notice that all the
functions multiplying the projectors depend only on {}{ the
parameters $c_{1,2}$ of the representations, while} all the other
parameters are hidden in the projectors.

\section{Conjugate representation and antiparticles}\label{section3}

The representation we consider in this section is the {\it
conjugate}, {\it i.e.} the supertranspose, representation of the
one studied in the previous section for the distinguished Dynkin
diagram. One can show that such representation is generated by

\begin{eqnarray}
\label{algebra2}
&&E_1 = E_{34}, \qquad F_1 = E_{43}, \qquad H_1 = [E_1,F_1],\\
&&E_2 = - b \, E_{23} - \, a E_{41}, \qquad F_2 = - e \, E_{14} -
d E_{32}, \qquad H_2 = [E_2,F_2],\nonumber
\end{eqnarray}
where the parameters are constrained as

\begin{eqnarray}
a e = c + 1, \qquad b d = c.
\end{eqnarray}
The vector space on which this representation acts is again
generated by two bosons (indices $1$ and $2$) and two fermions
(indices $3$ and $4$). The Cartan matrix is the same as in the
previous section. The following assignment

\begin{eqnarray}
\label{102}
&&\xi^+_{i,0} = E_i, \qquad \xi^-_{i,0} = F_i, \qquad \kappa_{i,0} = H_i,\nonumber\\
&&\xi^+_{1,1} = u \, E_1, \qquad \xi^-_{1,1} = u \, F_1, \qquad \kappa_{1,1} = u \, H_1,\nonumber\\
&&\xi^+_{2,1} = - b \, \Big(u - \frac{1}{2}\Big) \, E_{23} - \, a \, \Big(u + \frac{1}{2}\Big) E_{41}, \qquad \xi^-_{2,1} = - e \, \Big(u + \frac{1}{2}\Big) \, E_{14} - d \, \Big(u - \frac{1}{2}\Big) E_{32}, \nonumber\\
&&\kappa_{2,1} = [\xi^+_{2,1},\xi^-_{2,0}],
\end{eqnarray}
with the rest of the generators $\xi^\pm_{i,n}$, $\kappa_{i,n}$,
$n>1$, consistently obtained by iteration of (\ref{def:drinf2}),
defines another representation of the same Yangian in Drinfeld's
second realization. One can promote to arbitrary levels the
representation (\ref{102}) simply by assigning
\begin{eqnarray}
\label{1021}
&&\xi^+_{1,n} = u^n \, E_1, \qquad \xi^-_{1,n} = u^n \, F_1, \qquad \kappa_{1,n} = u^n \, H_1,\nonumber\\
&&\xi^+_{2,n} = - b \, \Big(u - \frac{1}{2}\Big)^n \, E_{23} - \, a \, \Big(u + \frac{1}{2}\Big)^n E_{41}, \qquad \xi^-_{2,n} = - e \, \Big(u + \frac{1}{2}\Big)^n \, E_{14} - d \, \Big(u - \frac{1}{2}\Big)^n E_{32}, \nonumber\\
&&\kappa_{2,n} = [\xi^+_{2,n},\xi^-_{2,0}].
\end{eqnarray}

The $R$-matrix related to this Yangian representation must again
satisfy

\begin{eqnarray}
\label{inv2} \Delta^{op} ({\mathfrak{J}}) \, R = R \, \Delta
({\mathfrak{J}})
\end{eqnarray}
for any generator $\mathfrak{J}$ of the Yangian. The coproduct is
trivial on the entire level $n=0$ Lie subalgebra of the Yangian.

We consider in this section both representations 1 and 2 to be the
conjugate representation (\ref{102}). By looking at
(\ref{algebra2}) and recalling that the total number of particles
is conserved, we immediately obtain the conservation {}{law for
the differences of the} numbers of bosons and fermions, exactly as
in the previous section. The R-matrix (choosing the same overall
normalization as in the previous section) can therefore again be
again parametrized by the same equations
(\ref{Rmatrice}),(\ref{Rmatrice1}).

The comultiplication becomes non-trivial as soon as we move to the
level one Yangian generators. If we define once again

\begin{eqnarray}
E_3 = [E_1, E_2], \qquad F_3 = [F_1, F_2].
\end{eqnarray}
then we can directly use the formulas (\ref{coprid}), which are
universal for any representation\footnote{We have checked that
also in this representation the coproducts provide a homomorphism
of the Yangian, namely, they respect the relations
(\ref{def:drinf2}).} . The result of imposing the invariance of
the R-matrix (\ref{inv2}) is given in the appendix.

\bigskip

{}{Let us now} construct the antiparticle representation of the
Yangian representation (\ref{algebra}), (\ref{10}). Such
antiparticle representation is defined as

\begin{eqnarray}
\label{antialgebra}
&&E_1 = E_{43}, \qquad F_1 = E_{34}, \qquad H_1 = - E_{33} + E_{44},\\
&&E_2 = - \bar{a} \, E_{14} + \bar{b} E_{32}, \qquad F_2 = -
\bar{d} \, E_{23} + \, \bar{e} E_{41}, \qquad H_2 = - \bar{c} \,
\mathbbmss{1} - E_{11} - E_{44},\nonumber
\end{eqnarray}
where the parameters are constrained as

\begin{eqnarray}
\label{costri2} \bar{a} \bar{e} = \bar{c} + 1, \qquad \bar{b}
\bar{d} = \bar{c}
\end{eqnarray}
and such that

\begin{eqnarray}
\label{conju} {\cal{S}}(\mathfrak{J}) \, = \, \mathfrak{C}^{-1} \,
\bar{\mathfrak{J}}^{st} \, \mathfrak{C}.
\end{eqnarray}
In (\ref{conju}), $\mathfrak{J}$ is any generator in the
representation (\ref{10}), $\bar{\mathfrak{J}}$ is any generator
in the representation (\ref{antialgebra}) and corresponding
Yangian, $\mathfrak{C}$ is a suitable charge conjugation matrix,
and $\cal{S}$ is the Yangian Hopf algebra antipode\footnote{We
remind that the antipode is defined on the whole Hopf algebra by
the relation $\mu \, ({\cal{S}} \otimes \mathbbmss{1}) \Delta =
\eta \, \epsilon$ (and $\mu \, (\mathbbmss{1} \otimes
{\cal{S}}^{-1}) \Delta = \eta \, \epsilon$ for invertible
antipodes) involving the multiplication $\mu$, the coproduct
$\Delta$, the counit $\epsilon$ and the unit $\eta$. The counit
$\epsilon$ turns out to act as zero on all generators of the
Yangian, while one has $\epsilon (\mathbbmss{1})=1$.}.

In order to find a solution to the condition (\ref{conju}), we
also need to allow the Yangian related to the representation
(\ref{antialgebra}) to have a spectral parameter $\bar{u}$
different from $u$ in (\ref{10}). If we do that, we can find a
consistent solution which reads

\begin{eqnarray}
\label{antipar1} \mathfrak{C} = \frac{1}{\bar{a}} \, b \, E_{12}
\, + \, \frac{1}{a} \, \bar{b} \, E_{21} \, - \, E_{34} \, + \,
E_{43},
\end{eqnarray}
\begin{eqnarray}
\label{antipar2} \bar{c} = - c - 1, \qquad \bar{u} = u + c.
\end{eqnarray}
Notice that the combination

\begin{eqnarray}
\vartheta = - 2 \pi i \, (u + \frac{c}{2}),\label{theta}
\end{eqnarray}
transforms as a relativistic rapidity under the crossing
transformation (\ref{antipar2}), {\it i.e.}

\begin{eqnarray}
\bar{\vartheta} \, = \, \vartheta \, + \, i \pi.
\end{eqnarray}
The R-matrix depends on the variables $\vartheta_1$ and
$\vartheta_2$ only through their difference, consistently with the
existence of a {\it shift} automorphism of the Yangian. The
representations we find in this paper are all of the so-called
{\it evaluation} type \cite{Ale1,Ale2}. In such representations,
the shift automorphism simply transforms the spectral parameter as
$u \to u + q$, where $q$ is a constant independent on the
representation.

We can also introduce a set of parameters which are reminiscent of
AdS/CFT \cite{Beisert:2005tm}. In fact, let us make the following
choice:

\begin{eqnarray}
\label{Janik} a = -1, \qquad b = -\alpha \bigg( 1 -
\frac{x^-}{x^+}\bigg), \qquad d = \frac{i \beta}{x^-}, \qquad e =
i (x^+ - x^-),
\end{eqnarray}
with

\begin{eqnarray}
x^+ + \frac{\alpha \, \beta}{x^+} - x^- - \frac{\alpha \,
\beta}{x^-} \, = \, i.
\end{eqnarray}
One can check that the constraint (\ref{costri}) is satisfied,
with

\begin{eqnarray}
c = - 1 - i (x^+ - x^-).
\end{eqnarray}
In terms of these new variables, the antiparticle transformation
(\ref{antipar2}) for the variable $c$ amounts to the same map
found by Janik \cite{Janik:2006dc} in the AdS/CFT
context\footnote{It is interesting to notice how the same map
arises in the AdS/CFT context from imposing the condition $\mu \,
({\cal{S}} \otimes \mathbbmss{1}) \Delta = \eta \, \epsilon$ on a
nontrivial level zero coproduct, as showed in \cite{PST}.}, namely

\begin{eqnarray}
\bar{x}^\pm \, = \, \frac{\alpha \, \beta}{x^\pm}.
\end{eqnarray}
This map can be then expressed in terms of a generalized rapidity
by means of Weierstrass functions (see \cite{Janik:2006dc}). We
also notice that, with the assignment (\ref{Janik}), the
representation (\ref{algebra2}) becomes precisely the AdS/CFT
representation used in \cite{Janik:2006dc}.

\section{Crossing symmetry and S matrix}\label{section4}

The mixed R-matrix which intertwines a representation of the type
(\ref{antialgebra}) with a representation of the type
(\ref{algebra}) has to satisfy the following crossing-symmetry
condition:

\begin{eqnarray}
\label{crossu} (\mathfrak{C}^{-1} \otimes \mathbbmss{1}) \,
\Phi_{\bar{1}2} \, R_{\bar{1}2}^{st_1} \, (\mathfrak{C} \otimes
\mathbbmss{1}) \, \Phi_{12} \, R_{12} \, = \, \mathbbmss{1}
\otimes \mathbbmss{1},
\end{eqnarray}
derived from the following condition one imposes on the universal
R-matrix (which is assumed to be invertible):

\begin{eqnarray}
\label{r1} ({\cal{S}} \otimes \mathbbmss{1}) \, R \, = \, R^{-1}.
\end{eqnarray}
In (\ref{crossu}), $st_1$ means taking the supertranspose in the
space $1$ of the tensor product, the charge conjugation matrix is
given by (\ref{antipar1}). The R-matrix $R_{12}$ coincides with
the one we have obtained in section \ref{section2}, while the
mixed R-matrix $R_{\bar{1}2}$ is given by straightforward
substitution of {}{the} representation $1$ with its associated
antiparticle representation. We report the result here below for
{}{the} convenience of the reader:

\begin{eqnarray}
\label{Rmatrice1bar2}
&&R_{\bar{1}2} \, {\mathfrak{11}} = {\mathfrak{11}},\nn\\
&&R_{\bar{1}2} \, {\mathfrak{12}} = B' \, {\mathfrak{12}} + C' \, {\mathfrak{21}} + D' \, {\mathfrak{34}} + E' \, {\mathfrak{43}},\nn\\
&&R_{\bar{1}2} \, {\mathfrak{21}} = F' \, {\mathfrak{12}} + G' \, {\mathfrak{21}} + H' \, {\mathfrak{34}} + I' \, {\mathfrak{43}},\nn\\
&&R_{\bar{1}2} \, {\mathfrak{22}} = L' \, {\mathfrak{22}},\nn\\
&&R_{\bar{1}2} \, {\mathfrak{33}} = \Gamma' \, {\mathfrak{33}},\nn\\
&&R_{\bar{1}2} \, {\mathfrak{34}} = P' \, {\mathfrak{12}} + Q' \, {\mathfrak{21}} + N' \, {\mathfrak{34}} + \Theta' \, {\mathfrak{43}},\nn\\
&&R_{\bar{1}2} \, {\mathfrak{43}} = T' \, {\mathfrak{12}} + U' \, {\mathfrak{21}} + \Psi' \, {\mathfrak{34}} + \Xi' \, {\mathfrak{43}},\nn\\
&&R_{\bar{1}2} \, {\mathfrak{44}} = V' \, {\mathfrak{44}},
\end{eqnarray}

\begin{eqnarray}
&&R_{\bar{1}2} \, {\mathfrak{13}} = \alpha_1' \, {\mathfrak{13}} + \alpha_2' \, {\mathfrak{31}},\nn\\
&&R_{\bar{1}2} \, {\mathfrak{14}} = \alpha_3' \, {\mathfrak{14}} + \alpha_4' \, {\mathfrak{41}},\nn\\
&&R_{\bar{1}2} \, {\mathfrak{23}} = \alpha_5' \, {\mathfrak{23}} + \alpha_6' \, {\mathfrak{32}},\nn\\
&&R_{\bar{1}2} \, {\mathfrak{24}} = \alpha_7' \, {\mathfrak{24}} + \alpha_8' \, {\mathfrak{42}},\nn\\
&&R_{\bar{1}2} \, {\mathfrak{31}} = \beta_1'\, {\mathfrak{13}} + \beta_2' \, {\mathfrak{31}},\nn\\
&&R_{\bar{1}2} \, {\mathfrak{41}} = \beta_3' \, {\mathfrak{14}} + \beta_4' \, {\mathfrak{41}},\nn\\
&&R_{\bar{1}2} \, {\mathfrak{32}} = \beta_5' \, {\mathfrak{23}} + \beta_6' \, {\mathfrak{32}},\nn\\
&&R_{\bar{1}2} \, {\mathfrak{42}} = \beta_7' \, {\mathfrak{24}} +
\beta_8' \, {\mathfrak{42}},
\end{eqnarray}

\begin{eqnarray}
\delta u = u_1 - u_2,
\end{eqnarray}

\begin{eqnarray}
&&B' = \frac{(\delta u - 1 - c_2)(\delta u - c_2)}{(- 1 + \delta u + c_1 - c_2)(\delta u + c_1 - c_2)}, \qquad C' = \frac{b_2 (1 + c_2) \bar{d}_1 \bar{e}_1}{(- 1 + \delta u + c_1 - c_2)(\delta u + c_1 - c_2) e_2},\nonumber\\
&&D' = - E' = - \frac{b_2 (\delta u - 1 - c_2) \bar{e}_1}{(- 1 + \delta u + c_1 - c_2)(\delta u + c_1 - c_2)}, \nonumber\\
&&F' = \frac{\bar{a}_1 \bar{b}_1 (1 + c_2) d_2}{a_2 (- 1 + \delta u + c_1 - c_2)(\delta u + c_1 - c_2)},\nonumber\\
&&G' = \frac{(\delta u + c_1)(1 + \delta u + c_1)}{(- 1 + \delta u + c_1 - c_2)(\delta u + c_1 - c_2)}, \qquad V' = \Gamma', \qquad T'=-P',\nonumber\\
&&H' = - I' = - \frac{(\delta u + c_1)\, \bar{b}_1 (1 + c_2)}{a_2
(- 1 + \delta u + c_1 - c_2)(\delta u + c_1 - c_2)}, \qquad \Xi' =
N',\nonumber
\end{eqnarray}
\begin{eqnarray}
&&L' = \frac{(\delta u - 1) \delta u}{(- 1 + \delta u + c_1 - c_2)(\delta u + c_1 - c_2)}, \qquad \Gamma' = \frac{\delta u}{(- 1 + \delta u + c_1 - c_2)}, \qquad \Psi' = \Theta', \nonumber\\
&&N' = \frac{(\delta u + c_1)(\delta u - 1 - c_2)}{(- 1 + \delta u + c_1 - c_2)(\delta u + c_1 - c_2)}, \qquad \Theta' = \frac{\delta u + c_1 (1 + c_2)}{(- 1 + \delta u + c_1 - c_2)(\delta u + c_1 - c_2)},\nonumber\\
&&P' = - \frac{\bar{a}_1 d_2 (\delta u - 1 - c_2)}{(- 1 + \delta u + c_1 - c_2)(\delta u + c_1 - c_2)}, \nonumber\\
&&Q' = - U' = - \frac{\bar{d}_1 \, (\delta u + c_1)\, (1 +
c_2)}{(- 1 + \delta u + c_1 - c_2)(\delta u + c_1 - c_2) e_2},
\end{eqnarray}

\begin{eqnarray}
&&\alpha_1' = \alpha_3' = \frac{\delta u - 1 - c_2}{- 1 + \delta u + c_1 - c_2}, \qquad \alpha_2' = \alpha_4' = - \frac{a_2 c_1}{\bar{a}_1 (1 - \delta u - c_1 + c_2)},\nonumber\\
&&\alpha_5' = \alpha_7' = \frac{\delta u (\delta u + c_1)}{(- 1 + \delta u + c_1 - c_2)(\delta u + c_1 - c_2)},\nonumber\\
&&\alpha_6' = \alpha_8' = \frac{\bar{b}_1 \, \delta u \, d_2}{(- 1 + \delta u + c_1 - c_2)(\delta u + c_1 - c_2)},\nonumber\\
&&\beta_1' = \beta_3' = - \frac{c_1 e_2}{(1 - \delta u - c_1 + c_2) \bar{e}_1}, \qquad \beta_2' = \beta_4' = \frac{\delta u + c_1}{- 1 + \delta u + c_1 - c_2},\nonumber\\
&&\beta_5' = \beta_7' = \frac{b_2 \, \delta u \, \bar{d}_1}{(- 1 + \delta u + c_1 - c_2)(\delta u + c_1 - c_2)}, \nonumber\\
&&\beta_6' = \beta_8' = \frac{\delta u \, (\delta u - 1 - c_2)}{(-
1 + \delta u + c_1 - c_2)(\delta u + c_1 - c_2)}.
\end{eqnarray}
Moreover, the overall scalar factor multiplying the R-matrix is

\begin{eqnarray}
\label{ov1} \Phi_{12} \, = \,
\Phi(a_1,b_1,c_1,d_1,e_1,u_1,a_2,b_2,c_2,d_2,e_2,u_2),
\end{eqnarray}
and
\begin{eqnarray}
\label{ov2} \Phi_{\bar{1}2} \, = \, \Phi(\bar{a}_1,\bar{b}_1,- c_1
- 1,\bar{d}_1,\bar{e}_1,u_1 + c_1,a_2,b_2,c_2,d_2,e_2,u_2),
\end{eqnarray}
where the scalar function $\Phi$ appearing in (\ref{ov1}),
(\ref{ov2}) is fixed by requiring crossing symmetry and unitarity
to hold. Notice also that, in the parametrization given by
(\ref{Janik}), one has

\begin{eqnarray}
\label{Janik2} \bar{a}_m = -1, \qquad \bar{b}_m = -\alpha \bigg( 1
- \frac{x_m^+}{x_m^-}\bigg), \qquad \bar{d}_m = \frac{i
x_m^-}{\alpha}, \qquad \bar{e}_m = - 1 - i (x_m^+ - x_m^-).
\end{eqnarray}
with $m=1,2$.

By making use of the above expressions, one can show that the
crossing condition (\ref{crossu}) reduces to the following
equation for the scalar factor $\Phi$:

\begin{eqnarray}
\label{crossi1} \Phi_{12} \, \Phi_{\bar{1}2} \, = \, \frac{(c_2 +
u_2 - u_1 - c_1)(1 + c_2 + u_2 - u_1 - c_1)}{(- u_2 + u_1 + c_1)(1
- u_2 + u_1 + c_1)} \equiv f(c_1,c_2,u_1,u_2).
\end{eqnarray}
In terms of the parameters $x^\pm$ (\ref{Janik}) and
(\ref{Janik2}) this reads

\begin{eqnarray}
&&\Phi (x_1^\pm,x_2^\pm,u_1,u_2) \, \Phi (\frac{\alpha \beta}{x_1^\pm},x_2^\pm,u_1 -1-i(x_1^+ - x_1^-),u_2)\\
&&\qquad = \, \frac{(u_2-u_1-i(x_1^- - x_2^- - x_1^+ +
x_2^+))(u_2-u_1-i(i+x_1^- - x_2^- - x_1^+ +
x_2^+))}{(u_2-u_1-i(x_1^- - x_1^+))(1+u_2-u_1-i x_1^- +i
x_1^+))}.\nonumber
\end{eqnarray}

It is quite interesting to notice what happens when considering
antiparticles in the second factor of the tensor product. In fact,
the condition on the universal R-matrix complementary to
(\ref{r1}) is (for an invertible antipode map)

\begin{eqnarray}
(\mathbbmss{1} \otimes {\cal{S}}^{-1}) \, R \, = \, R^{-1},
\end{eqnarray}
which means that in the second factor of the tensor product we
have to analyze the equation
\begin{eqnarray}
\label{conju2} {\cal{S}}^{-1} \, (\mathfrak{J}) \, = \,
\tilde{\mathfrak{C}}^{-1} \, \tilde{\mathfrak{J}}^{st} \,
\tilde{\mathfrak{C}},
\end{eqnarray}
complementary to (\ref{conju}). In order to do this, we notice
that ${\cal{S}}^2 = \mathbbmss{1}$ on the level zero of the
Yangian (since at level zero the antipode just changes the sign to
any generator). This means that {}{the} inverse of the antipode
equals the antipode itself at level zero, and the condition
(\ref{conju2}) coincides with (\ref{conju}) at level zero, which
fixes

\begin{eqnarray}
\label{lde} \tilde{\mathfrak{C}} \, = \, \mathfrak{C}, \qquad
\tilde{c} = \bar{c} = - c - 1.
\end{eqnarray}
At Yangian-level instead, one has the following:

\begin{eqnarray}
{\cal{S}}^2 \, (\xi^\pm_{i,1}) = \xi^\pm_{i,1} \, - \,
\xi^\pm_{i,0}.
\end{eqnarray}
This means

\begin{eqnarray}
{\cal{S}}^{-1} \, (\xi^\pm_{i,1}) = {\cal{S}} (\xi^\pm_{i,1}) \, -
\,  \xi^\pm_{i,0},
\end{eqnarray}
which in turn implies that (\ref{conju2}) is solved for all
generators by the following requirement:

\begin{eqnarray}
\label{crossi2} \tilde{u} \, = \, \bar{u} + 1 \, = \, u + c + 1.
\end{eqnarray}
We then consider the R-matrix, and indeed we find that it
satisfies the analog of Eq. (\ref{crossu}), this time for the
inverse-antipodal representation we just found, namely

\begin{eqnarray}
\label{crossu2} (\mathbbmss{1} \otimes \mathfrak{C}^{-1}) \,
\Phi_{1\tilde{2}} \, R_{1\tilde{2}}^{st_2} \, (\mathbbmss{1}
\otimes \mathfrak{C}) \, \Phi_{12} \, R_{12} \, = \, \mathbbmss{1}
\otimes \mathbbmss{1}.
\end{eqnarray}
The R-matrix $R_{1\tilde{2}}$ is obtained by substituting the
antiparticle representation (\ref{lde}), (\ref{crossi2}) in the
second factor of the tensor product. One can show that the
relation (\ref{crossu2}) amounts to the following requirement for
the overall scalar factor $\Phi$ of (\ref{ov1}):

\begin{eqnarray}
\label{crossuu2} \Phi_{12} \, \Phi_{1\tilde{2}} \, = \, \Bigg[1 +
\frac{c_1(1 + c_1)}{u_2 - u_1} - \frac{(2 + c_1)(1 + c_1)}{u_2 -
u_1 +1}\Bigg]^{-1} \equiv g(c_1,c_2,u_1,u_2).
\end{eqnarray}
with
\begin{eqnarray}
\label{ovv2} \Phi_{1\tilde{2}} \, = \,
\Phi(a_1,b_1,c_1,d_1,e_1,u_1,\tilde{a}_2,\tilde{b}_2,- c_2 -
1,\tilde{d}_2,\tilde{e}_2,u_2 + c_2 + 1).\nonumber
\end{eqnarray}

Let us comment on consistency of the crossing relation, double
crossing and unitarity. First, if we apply the crossing
transformation (\ref{antipar2}) on particle $1$ one more time to
(\ref{crossi1}), we schematically obtain (explicitly displaying
only the variables affected by the transformation)

\begin{eqnarray}
\Phi(-c_1-1,u_1+c_1) \, \Phi(c_1, u_1 - 1) =
f(-c_1-1,c_2,u_1+c_1,u_2),
\end{eqnarray}
with no apparent contradiction with (\ref{crossi1}). Similarly,
applying one more time the crossing (\ref{crossi2}) on particle
$2$ to (\ref{crossuu2}) results in

\begin{eqnarray}
\Phi(-c_2-1,u_2+c_2+1) \, \Phi(c_2, u_2 + 1) = \,
g(c_1,-c_2-1,u_1,u_2+c_2+1),
\end{eqnarray}
with no apparent contradiction with (\ref{crossuu2}). Finally, if
we consider the unitarity relation (\ref{unita}) and the two
crossing relations (\ref{crossi1}) and (\ref{crossuu2}), we can
deduce both

\begin{eqnarray}
\label{Cnst1} \Phi_{12}^{-1} \, \Phi_{\bar{1}2}^{-1} \, = \,
\Phi_{21} \, \Phi_{2\bar{1}} \, = f^{-1} (c_1,c_2,u_1,u_2)
\end{eqnarray}
and at the same time
\begin{eqnarray}
\label{Cnst2} \Phi_{21} \, \Phi_{2\bar{1}} \, =
g(c_2,-c_1-1,u_2,u_1+c_1).
\end{eqnarray}
The latter formula is obtained by exchanging $1$ and $2$ in
(\ref{crossuu2}), and subsequently sending $c_1 \to - c_1 - 1$ and
$u_1 \to u_1 + c_1$, in such a way that $\Phi_{21} \,
\Phi_{2\tilde{1}} = \Phi(c_1,u_1) \, \Phi(-c_1-1, u_1 +c_1+1)$
precisely becomes $\Phi_{21} \, \Phi_{2\bar{1}} = \Phi(c_1,u_1) \,
\Phi(-c_1-1, u_1 +c_1)$. By taking into account the explicit form
of the functions $f$ and $g$, one can check that (\ref{Cnst1}) and
(\ref{Cnst2}) are consistent with each other.

Notice that we can find a solution to (\ref{crossi1}) and
(\ref{crossuu2}) simultaneously, namely

\begin{eqnarray}
\Phi^{(0)}_{12} = \frac{\Gamma(1 + c_2 - u_1 + u_2) \, \Gamma(2 +
c_2 - u_1 + u_2)\Gamma(-1 - c_1 - u_1 + u_2)\Gamma(- c_1 - u_1 +
u_2)}{\Gamma(- u_1 + u_2) \, \Gamma(1 - u_1 + u_2)\Gamma(c_2 - c_1
- u_1 + u_2)\Gamma(1+c_2- c_1 - u_1 + u_2)}\nonumber.
\end{eqnarray}
However, the above factor is not unitary. In fact, solving
(\ref{crossi1}) and (\ref{crossuu2}) simultaneously only implies
for instance

\begin{eqnarray}
\Phi^{(0)}_{12} \, \Phi^{(0)}_{21} \, \Phi^{(0)}_{\bar{1}2} \,
\Phi^{(0)}_{2\bar{1}} \, = \, 1,
\end{eqnarray}
which is not equivalent to the relation (\ref{unita}) (although it
is compatible with it). A formal solution of (\ref{crossi1}) and
(\ref{crossuu2}) which is also unitary is then obtained as

\begin{eqnarray}
\label{tor} \Phi_{12} =
\sqrt{\frac{\Phi^{(0)}_{12}}{\Phi^{(0)}_{21}}}.
\end{eqnarray}

As a further check, we have computed the scalar factor as it comes
from evaluating the universal R-matrix \cite{KT,RS} on the
all-level Yangian representation (\ref{10f}), and found that,
after unitarization, it precisely coincides with (\ref{tor}).
{}{More precisely, the universal R-matrix reads
\begin{eqnarray}
\label{univ} &&{ R}={ R}_E { R}_H { R}_F,
\end{eqnarray}
where {\rm $R_E$ and $R_F$ are certain factors depending on the
Yangian generators associated to the positive and negative roots
of the algebra, and
\begin{eqnarray}
&&{ R}_H= \exp \left\{ {\rm
Res}_{u=v}\left[\sum_{i,j} \frac{\rm d}{{\rm d}u}({ \log
}H_i^+(u))\otimes { } D^{-1}_{ij} \log
H_j^-(v)\right]\right\},
\end{eqnarray}
where $D_{ij} = - (T^{\frac{1}{2}} - T^{-\frac{1}{2}}) \, a_{ij}
(T^{\frac{1}{2}})$, $a_{ij} (q) = \frac{q^{a_{ij}} - \,
q^{-a_{ij}}}{q - q^{-1}}$ with $a_{ij}$ the Cartan matrix entries,
and the operator $T$ is defined such that $T f(u) = f(u+1)$. One
also defines
\begin{eqnarray}
\label{eqn;Res} &&{\rm Res}_{u=v}\left(A(u)\otimes
B(v)\right)=\sum_k a_k\otimes b_{-k-1}
\end{eqnarray}
for $A(u)=\sum_k a_k u^{-k-1}$ and $B(u)=\sum_k b_k u^{-k-1}$, and
the so-called {\it Drinfeld's currents} (for the Cartan
subalgebra) are given by
\begin{eqnarray}
\label{curr} &&H_i^{\pm}(u)=1\pm \sum_{n \ge 0 \atop n<0} \kappa_{i,n}
u^{-n-1} ~.~~~~
\end{eqnarray}
In order to determine the scalar factor, we can simply act on the
state $\mathfrak{11}$. The root factors $R_E$ and $R_F$ act as
identity, and all one is left with is calculating the contribution
from the Cartan part $R_H$. We adopt the prescription of \cite{RS}
and everywhere interpret
\begin{eqnarray}
\frac{1}{T^{\frac{1}{2}} - T^{-\frac{1}{2}}} = - \sum_{p=0}^\infty
T^{p + \frac{1}{2}}.
\end{eqnarray}
A tedious calculation utilizing the procedure in Appendix A.2 of
\cite{Arutyunov:2009ce} gives}
\begin{eqnarray}
R_H \, \mathfrak{11} \, = \, \frac{\Gamma(u_1 - u_2) \Gamma(1 +
u_1 - u_2) \Gamma(c_1 - c_2 + u_1 - u_2) \Gamma(
  1 + c_1 - c_2 + u_1 - u_2) \, \mathfrak{11}}{
\Gamma(1 + c_1 + u_1 - u_2) \Gamma(
  2 + c_1 + u_1 - u_2) \Gamma(-1 - c_2 + u_1 - u_2) \Gamma(-c_2 + u_1 - u_2)}.\nonumber
\end{eqnarray}
Unitarizing this result in the fashion (\ref{tor}) produces a
scalar factor which coincides with what is obtained by unitarizing
$\Phi^{(0)}_{12}$.}

We finish by noticing that the R-matrix $R_{\bar{1}\tilde{2}}$ can
easily be obtained by substituting the appropriate representations
in the two tensor product factors. Furthermore, in order to obtain
the physical S-matrix one needs to apply the graded permutation
operator to any R-matrix from this paper, {\it i.e.} $S= PR$.

It is convenient to write down the {}{R}-matrix {}{$R_{12}$},
including the crossing-unitarity factor, in terms of the variable
$\vartheta$ (\ref{theta}). We define
$x=(\vartheta_2-\vartheta_1)/\pi i$, in terms of which the unitary
and crossing symmetric S-matrix reads
\begin{eqnarray}
\label{tot} {}{R}_{12}(x)&=& \left[ \frac{\Gamma
(1-x/2+\tilde{c})\Gamma(2-x/2+\tilde{c})
\Gamma(-1-x/2-\tilde{c})\Gamma(-x/2-\tilde{c})}
{\Gamma(-x/2-\delta c)\Gamma(1-x/2-\delta c)\Gamma(-x/2+\delta c)
\Gamma(1-x/2+\delta c)}\right. \times \nonumber\\
&&\left.\frac{\Gamma(x/2+\delta c)\Gamma(1+x/2+\delta c)
\Gamma(x/2-\delta c)\Gamma(1+x/2-\delta c)}
{\Gamma(1+x/2+\tilde{c})\Gamma(2+x/2+\tilde{c})\Gamma(-1+x/2-\tilde{c})\Gamma(x/2-\tilde{c})}
\right]^{1/2} \times \nonumber\\
&&\left\{ P_2 + \frac{(x/2+\tilde{c})(x/2+\tilde{c}+1)}
{(x/2-\tilde{c})(x/2-\tilde{c}-1)} P_1 +
\frac{x/2+\tilde{c}+1}{x/2-\tilde{c}-1} P_3 \right\} \label{Sm}.
\end{eqnarray}
Here we have defined
\begin{eqnarray}
\tilde{c} = \frac{c_1+c_2}{2}, ~ ~\delta c = \frac{c_2-c_1}{2}.
\nonumber
\end{eqnarray}

{}{In appendix B we report the structure of poles of the R-matrix
in the physical sheet $0<x<1$.}

\section{Conclusions}

In this paper we have {\rm adopted a Yangian construction to
independently rederive\footnote{\label{f2} \rm As we pointed out,
there most likely exists a limit where the quantum affine result
$R^{V_4 \, V_4'}$ in \cite{Gru2} reduces to R-matrices similar to
ours.}} a rational R-matrix with $\mathfrak{sl}(2|1)$ Yangian
symmetry in a four dimensional representation and its conjugate
(antiparticle). Each of these representations depend on three
additional continuum parameters. The calculation was done by
working out of explicit form of the Yangian representation in the
so-called Drinfeld's second realization, and by making use of the
associated Hopf-algebra coproducts. {\rm Our methodology allows
us} to interpret the found R-matrix as a relativistic scattering
S-matrix. {\rm With this we mean that we have derived consistent unitarity
and crossing relations for this R-matrix and the associated Yangian representations. We have then solved these relations}, determining in
this way the overall scalar factor of the R-matrix (apart from
possible CDD factors). The scalar factor we single out corresponds
to the unitarization of the scalar factor coming from the
universal R-matrix.

Let us point out further steps of investigation. {\rm We plan to utilize the results we have obtained in this paper} as a starting point for the
investigation of the spectrum of integrable alternating spin
chains with $\mathfrak{sl}(2|1)$ symmetry in four dimensional
representations, their thermodynamics and their conformal limit
spectrum. Especially interesting are the questions about the
dependence of thermodynamic and conformal properties of the spin
chain on the continuum parameters of the four dimensional
representation, and most crucially about the effect of the
dressing phase we have derived on the spectral properties of the
theory.

One of the possible physical interpretations of the obtained
S-matrix can be found in the context of SSSG model with the
$\mathfrak{osp}(3|2)/\mathfrak{osp}(2|2)$ symmetry. A check of
such correspondence can be attempted by using thermodynamic Bethe
ansatz techniques based on the Bethe equations. To this purpose,
an important step will be the investigation of the bound state
spectrum encoded in the poles of the S-matrix in the physical
strip.

Finally, it is very {}{interesting} to notice how the R-matrices
and representations we have obtained are very similar to the ones
one encounters in context of AdS/CFT integrability, and the
comparison can be very fruitful in terms of a better understanding
of the features of the AdS/CFT Yangian in Drinfeld's second
realization for various choices of the Dynkin diagram.

We hope to return to this and other questions in further
publications.

\section{Acknowledgements}
A. Babichenko is thankful to the Einstein center of Weizmann
Institute for support. A. Torrielli thanks the UK EPSRC for
funding under grant EP/H000054/1 during the initial stage of this
work, and Nordita-Stockholm for hospitality during a
{}{subsequent} stage of this work.

\section{Appendix A}

We report here below the R-matrix intertwining two conjugate
representations of section \ref{section3}. Defining
\begin{eqnarray}
\delta u = u_1 - u_2,
\end{eqnarray}
one finds

\begin{eqnarray}
&&B = \frac{\delta u (1 + \delta u)}{(- 1 + \delta u - c_1)(\delta u - c_1)}, \qquad C = \frac{a_1 b_1 c_2 e_2}{(- 1 + \delta u - c_1)(\delta u - c_1) b_2}, \qquad T = -P\nonumber\\
&&D = - E = \frac{\delta u \, a_1 c_2}{b_2 (- 1 + \delta u - c_1)(\delta u - c_1)}, \qquad F = \frac{a_2 b_2 (1 + c_1) c_1}{a_1 b_1 (- 1 + \delta u - c_1)(\delta u - c_1)},\nonumber\\
&&G = \frac{(\delta u - c_1 + c_2)(1 + \delta u - c_1 + c_2)}{(- 1 + \delta u - c_1)(\delta u - c_1)}, \qquad H = - I = \frac{a_2 d_1 (\delta u - c_1 + c_2)}{(- 1 + \delta u - c_1)(\delta u - c_1)}, \nonumber\\
&&L = \frac{(\delta u + c_2) (1 + \delta u + c_2)}{(- 1 + \delta u - c_1)(\delta u - c_1)}, \qquad \Gamma = \frac{(1 + \delta u + c_2)}{(- 1 + \delta u - c_1)}, \qquad \Psi = \Theta, \qquad \Xi = N,\nonumber\\
&&N = \frac{\delta u (\delta u - c_1 + c_2)}{(- 1 + \delta u - c_1)(\delta u - c_1)}, \qquad \Theta = \frac{\delta u - c_1 (1 + c_2)}{(- 1 + \delta u - c_1)(\delta u - c_1)}, \qquad V = \Gamma,\nonumber\\
&&P = \frac{\delta u \, b_2 (1 + c_1)}{a_1 (- 1 + \delta u -
c_1)(\delta u - c_1)}, \qquad Q =  - U = \frac{b_1 e_2 (\delta u -
c_1 + c_2)}{(- 1 + \delta u - c_1)(\delta u - c_1)},
\end{eqnarray}
\begin{eqnarray}
&&\alpha_1 = \alpha_3 = \frac{\delta u}{- 1 + \delta u - c_1}, \qquad \alpha_2 = \alpha_4 =  \frac{a_1 e_2}{(1 - \delta u + c_1)},\nonumber\\
&&\alpha_5 = \alpha_7 = \frac{(1 + \delta u + c_2)(\delta u - c_1 + c_2)}{(- 1 + \delta u - c_1)(\delta u - c_1)}, \qquad \alpha_6 = \alpha_8 = \frac{c_1 b_2 (1 + \delta u + c_2)}{b_1 (- 1 + \delta u - c_1)(\delta u - c_1)},\nonumber\\
&&\beta_1 = \beta_3 = \frac{(1 + c_1) a_2}{(1 - \delta u + c_1) a_1}, \qquad \beta_2 = \beta_4 = \frac{\delta u - c_1 + c_2}{- 1 + \delta u - c_1},\\
&&\beta_5 = \beta_7 = \frac{b_1 d_2 (1 + \delta u + c_2)}{(- 1 +
\delta u - c_1)(\delta u - c_1)}, \qquad \beta_6 = \beta_8 =
\frac{(1 + \delta u + c_2) \, \delta u}{(- 1 + \delta u -
c_1)(\delta u - c_1)}\nonumber .
\end{eqnarray}

We have checked that this R-matrix satisfies the Yang-Baxter
equation.

\section{Appendix B}

Below we analyze the structure of singularities {}{of the R-matrix
(\ref{tot}) in the physical strip $0<x<1$}. First, we start by
listing the poles and zeroes of the the scalar factor. One can see
that most of the poles and zeroes under the square root are double
poles and zeroes, and those poles which are not  actually cancel
out, such that the remaining poles and zeroes after taking the
square root are all simple ones, and we are not left with square
root branch cuts.

The set of potential poles of the scalar factor\footnote{A
potential pole is such that it becomes a physical pole if it
belongs to the physical strip.} is
\begin{eqnarray}
\{2\delta c-2n\}~\bigcup~\{-2\delta c-2n\}~\bigcup~
\{2\tilde{c}+2n+2\}~\bigcup~\{-2\tilde{c}+2n-2\}, ~ ~
n=1,2,...\label{pp2}
\end{eqnarray}
and the set of zeroes is
\begin{eqnarray}
\{-2\delta c+2n\}~\bigcup~\{2\delta c+2n\}~\bigcup~
\{-2\tilde{c}-2n-2\}~\bigcup~\{2\tilde{c}-2n+2\}, ~ ~
n=1,2,...\label{zp2}
\end{eqnarray}

The analysis of potential poles and their cancellation with zeroes
in the physical strip leads to the following result, which we
describe in each scattering channel $P_{1,2,3}$ separately. {}{In
what follows, by $[a]$ (respectively, $\{a\}$) we mean the integer
(respectively, fractional) part of $a$}. For convenience we also
define the following functions:
\begin{eqnarray}
\label{ms}
m_1(a)&=& \min(-[a],[a]+2,0), ~~ m_2(a)=\max(-[a]-3,[a]-1,1),\nonumber\\
m_3(a)&=& \min(-[a],[a]+2,-1), ~~ m_4(a)=\max(-[a]-3,[a]-1,0),\nonumber\\
m_5(a)&=&\max(-[a]-5,[a]-1,0), ~~ m_6(a)= \min(-[a],[a]+2,-2)
\end{eqnarray}

$\bf{P_2}$ \textbf{channel}\\

The poles of the $P_2$ channel are defined {}{purely by the scalar
factor}. For generic values of $c_1$ and $c_2$, i.e. when neither
of them is integer, {}{nor} their sum or difference, there are
poles in the physical strip
\begin{itemize}
\item at $\{c_2-c_1\}$ if $c_2-c_1>2$ and $[c_2-c_1]$ is even
\item at $1-\{c_2-c_1\}$ if $c_2-c_1<-2$ and $[c_2-c_1]$ is odd
\item at $\{c_2+c_1\}$ if $c_2+c_1<-3$ and $[c_2+c_1]$ is even
\item at $1-\{c_2+c_1\}$ if $c_2+c_1>-1$ and $[c_2+c_1]$ is odd
\end{itemize}
 Notice that there are no poles if
$-5/2\leq c_1,c_2 \leq 1/2$. {}{As soon as $\{c_2-c_1\}$
(respectively, $\{c_2+c_1\}$) becomes integer, the poles at
$\{c_2-c_1\}$ and $1-\{c_2-c_1\}$ (respectively, $\{c_2+c_1\}$ and
$1-\{c_2+c_1\}$) fall out of the physical sheet (see below for a
remark about the special cases $\{c_2+c_1\}=0,-1,-2$).}

The picture becomes more complicated in the case when either $c_1$
or $c_2$ is integer{}{, but not both simultaneously}. If $c_{1}$
is integer but $c_{2}$ is not, the {} picture of poles in the
physical strip will be modified as follows.

\begin{itemize}
\item the pole at $\{c_2-c_1\}$ {}{can exist} if $[c_{2}-c_{1}]$ is even {}{ and $c_2-c_1>2$},
and it is a double pole coinciding with the pole at $\{c_2+c_1\}$
if $c_1<m_6(c_2)$, {}{it} is a simple pole if $m_6(c_2)\leq c_1 <
-[c_2]$, and it is {} cancelled by {}{a} zero if $-[c_2]\leq c_1$

\item the pole at $1-\{c_{2}-c_{1}\}$ {}{can exist} if $[c_{2}-c_{1}]$ is odd {}{ and $c_2-c_1<-2$},
and it is a double pole coinciding with the pole at
$1-\{c_{2}+c_{1}\}$ if $m_5(c_2)<c_1$, it is a simple pole if
$-[c_2]-5<c_1\leq m_5(c_2)$, and it is {} cancelled by {}{a} zero
if $c_1\leq -[c_2]-5$

\item the pole at $\{c_{2}+c_{1}\}$ {}{can exist} if $[c_{2}+c_{1}]$ is even {}{ and $c_2+c_1<-3$},
it is a double pole coinciding, as we said above, with the pole at
$\{c_2-c_1\}$ if $c_1<m_6(c_2)$, it is a simple pole if
$m_6(c_2)\leq c_1<[c_2]+2$, and it is {} cancelled by {}{a} zero
if $[c_2]+2\leq c_1$

\item the pole at $1-\{c_{2}+c_{1}\}$ {}{can exist} if $[c_{2}+c_{1}]$ is odd {}{ and $c_2+c_1>-1$},
it is a double pole coinciding, as we said above, with the pole at
$1-\{c_{2}-c_{1}\}$ if $m_5(c_2)<c_1$, it is a simple pole if
$[c_2]-1<c_1\leq m_5(c_2)$, and {}{ it is {}} cancelled by a zero
if $c_1\leq [c_2]-1$
\end{itemize}

If $c_{2}$ is integer but $c_{1}$ is not:

\begin{itemize}
\item the pole at $\{c_{2}-c_{1}\}$ {}{can exist} if $[c_{2}-c_{1}]$ is even {}{ and $c_2-c_1>2$},
it is a double pole coinciding with the pole at
$1-\{c_{2}+c_{1}\}$ if $m_5(c_1)<c_2$, it is a simple pole if
$-5-[c_1]<c_2\leq {}{m_5}(c_1)$, and {} {}{it is} cancelled by
{}{a} zero if $c_2\leq -5-[c_1]$

\item the pole at $1-\{c_{2}-c_{1}\}$ {}{can exist} if $[c_{2}-c_{1}]$ is
odd {}{ and $c_2-c_1<-2$}, it is a double pole coinciding with the
pole at $\{c_2+c_1\}$ if $c_2<m_6(c_1)$, it is a simple pole if
$m_6(c_1)\leq c_2<-[c_1]$, and it is {} cancelled by {}{a} zero if
$-[c_1]\leq c_2$

\item the pole at $\{c_{2}+c_{1}\}$ {}{can exist} if $[c_{2}+c_{1}]$ is
even {}{ and $c_2+c_1<-3$}, as we said, it is a double pole
coinciding with the pole at $1-\{c_{2}-c_{1}\}$ if $c_2<m_6(c_1)$,
it is a simple pole if $m_6(c_1)\leq c_2<[c_1]+2$, and it is {}
cancelled by {}{a} zero if $[c_1]+2\leq c_2$

\item the pole at $1-\{c_{2}+c_{1}\}$ {}{can exist} if $[c_{2}+c_{1}]$ is odd {}{ and $c_2+c_1>-1$},
it becomes a double pole coinciding, as we said above{}{,} with
the pole at $\{c_{2}-c_{1}\}$ if $m_5(c_1)<c_2$, it is a simple
pole if $[c_1]-1<c_2\leq m_5(c_1)$, and it is {} cancelled by
{}{a} zero if $c_2\leq [c_1]-1$
\end{itemize}

The spectral decomposition of channels $P_1$ and $P_3$ {}{is only slightly different from the one of} of the $P_2$ channel.\\

$\bf{P_1}$ \textbf{channel}\\

The factor before the projector $P_1$ cancels one pole in the set
of poles of the {}{scalar factor}, and adds to {}{the set} two
additional poles and one zero. For generic values of $c_1$ and
$c_2$ (see {}{remarks} above), in the $P_1$ channel there are
physical strip poles

\begin{itemize}
\item at $\{c_2-c_1\}$ if $c_2-c_1>2$ and $[c_2-c_1]$ is even
\item at $1-\{c_2-c_1\}$ if $c_2-c_1<-2$ and $[c_2-c_1]$ is odd
\item at $\{c_2+c_1\}$ if $c_2+c_1<{}{1}$ and $[c_2+c_1]$ is even
\item at $1-\{c_2+c_1\}$ if $c_2+c_1>1$ and $[c_2+c_1]$ is odd
\end{itemize}

If $c_1$ is integer, but not $c_{2}$, the picture of poles in the
$P_1$ channel {}{in the physical strip} is the following:

\begin{itemize}
\item the pole at $\{c_{2}-c_{1}\}$ {}{can exist} if $[c_{2}-c_{1}]$ is even {}{ and $c_2-c_1>2$},
it is a double pole coinciding with the pole at $\{c_{2}+c_{1}\}$
if $c_1<m_1(c_2)$, it is a simple pole if $m_1(c_2)\leq
c_1<-[c_2]$, and it is {} cancelled by {}{a} zero if $-[c_2]\leq
c_1$

\item the pole at $1-\{c_{2}-c_{1}\}$ {}{can exist} if $[c_{2}-c_{1}]$ is odd {}{ and $c_2-c_1<-2$},
it is a double pole coinciding with the pole at
$1-\{c_{2}+c_{1}\}$ if $m_2(c_2)<c_1$, it is a simple pole if
$-[c_2]-3<c_1\leq m_2(c_2)$, and it is {} cancelled by {}{a} zero
if $c_1\leq -[c_2]-3$

\item the pole at $\{c_{2}+c_{1}\}$ {}{can exist} if $[c_{2}+c_{1}]$ is even {}{ and $c_2+c_1<{}{1}$},
it is a double pole coinciding, as we said, with the pole at
$\{c_{2}-c_{1}\}$ if $c_1<m_1(c_2)$, it is a simple pole if
$m_1(c_2)\leq c_1<[c_2]+2$, and it is{} cancelled by {}{a} zero if
$[c_2]+2\leq c_1$

\item the pole at $1-\{c_{2}+c_{1}\}$ {}{can exist} if $[c_{2}+c_{1}]$ is odd {}{ and $c_2+c_1>1$},
it is a double pole coinciding, as we said, with the pole at
$1-\{c_{2}-c_{1}\}$ if $m_2(c_2)<c_1$, it is a simple pole if
$[c_2]-1<c_1\leq m_2(c_2)$, and it is {} cancelled by {}{a} zero
if $c_1\leq [c_2]-1$
\end{itemize}

{}{Conversely, if $c_2$ is integer, but $c_{1}$ is not,} the
situation is the following:

\begin{itemize}
\item the pole at $\{c_{2}-c_{1}\}$ {}{can exist} if $[c_{2}-c_{1}]$ is even {}{ and $c_2-c_1>2$},
it is a double pole coinciding with the pole at $1-\{c_2+c_1\}$ if
$m_2(c_1)<c_2$, it is a simple pole if $-[c_1]-3<c_2\leq
m_2(c_1)$, and it is{} cancelled by {}{a} zero if $c_2\leq
-[c_1]-3$

\item the pole at $1-\{c_{2}-c_{1}\}$ {}{can exist} if $[c_{2}-c_{1}]$ is
odd {}{ and $c_2-c_1<-2$}, it is a double pole coinciding with the
pole at ${}{\{c_{2}+c_{1}\}}$ if $c_2<m_1(c_1)$, it is a simple
pole if $m_1(c_1)\leq c_2<-[c_1]$, and it is {} cancelled by {}{a}
zero if $-[c_1]\leq c_2$

\item the pole at $\{c_{2}+c_{1}\}$ {}{can exist} if $[c_{2}+c_{1}]$ is even {}{ and $c_2+c_1<{}{1}$},
it is a double pole coinciding, as we said in the previous item,
with the pole at $1-\{c_{2}-c_{1}\}$ if $ c_2<m_1(c_1)$, it is a
simple pole if $m_1(c_1)\leq c_2<[c_1]+2$, and it is {} cancelled
by {}{a} zero if $[c_1]+2\leq c_2$

\item the pole at $1-\{c_{2}+c_{1}\}$ {}{can exist} if $[c_{2}+c_{1}]$ is odd {}{ and $c_2+c_1>1$},
it is a double pole coinciding with the pole at $\{c_{2}-c_{1}\}$
if $m_2(c_1)<c_2$, it is a simple pole if $[c_1]-1<c_2\leq
m_2(c_1)$, and it is {} cancelled by {}{a} zero if $c_2\leq
[c_1]-1$
\end{itemize}

$\bf{P_3}$ \textbf{channel}\\

For generic values of $c_1$ and $c_2$, the physical strip poles
are
\begin{itemize}
\item at $\{c_2-c_1\}$ if $c_2-c_1>2$ and $[c_2-c_1]$ is even
\item at $1-\{c_2-c_1\}$ if $c_2-c_1<-2$ and $[c_2-c_1]$ is odd
\item at $\{c_2+c_1\}$ if $c_2+c_1<-1$ and $[c_2+c_1]$ is even
\item at $1-\{c_2+c_1\}$ if $c_2+c_1>-1$ and $[c_2+c_1]$ is odd
\end{itemize}

For integer $c_1$ and not integer $c_2$ {}{their structure in the
physical strip is} modified as follows:

\begin{itemize}
\item the pole at $\{c_{2}-c_{1}\}$ {}{can exist} if $[c_{2}-c_{1}]$ is even {}{ and $c_2-c_1>2$},
and it is a double pole coinciding with the pole at $\{c_2+c_1\}$
if $c_1<m_3(c_2)$, it is a simple pole if $m_3(c_2)\leq
c_1<-[c_2]$, and it is {} cancelled by {}{a} zero if $-[c_2]\leq
c_1$

\item the pole at $1-\{c_{2}-c_{1}\}$ {}{can exist} if $[c_{2}-c_{1}]$ is
odd {}{ and $c_2-c_1<-2$}. The situation here is different for
$[c_2]<-3$ and for $[c_2]\geq -3$. If $[c_2]\geq-3$ it is a double
pole coinciding with the pole at $1-\{c_{2}+c_{1}\}$ if
$m_5(c2)<c_1$, and it is a simple pole if $-[c_2]-3<c_1\leq
m_5(c_2)$. If $[c_2]< -3$ it is a double pole coinciding with the
pole at $1-\{c_{2}+c_{1}\}$ if $-[c_2]-3<c_1$, and it is a simple
pole if $m_5(c_2)<c_1\leq -[c_2]-3$

\item the pole at $\{c_{2}+c_{1}\}$ {}{can exist} if $[c_{2}+c_{1}]$ is even {}{ and $c_2+c_1<-1$},
it is a double pole coinciding with the pole at $\{c_{2}-c_{1}\}$
if $c_1<m_3(c_2)$, it is a simple pole if $m_3(c_2)\leq
c_1<[c_2]+2$, and it is {} cancelled by {}{a} zero if $[c_2]+2\leq
c_1$

\item the pole at $1-\{c_{2}+c_{1}\}$ {}{can exist} if $[c_{2}+c_{1}]$ is odd {}{ and $c_2+c_1>-1$},
it is a double pole coinciding with the pole at
$1-\{c_{2}-c_{1}\}$ if $m_5(c_2)<c_1$, it is a simple pole if
$[c_2]-1<c_1\leq m_5(c_2)$, and it is {} cancelled by {}{a} zero
if $c_1\leq [c_2]-1$
\end{itemize}

{}{Conversely, if} $c_{2}$ is integer and $c_{1}$ is not

\begin{itemize}
\item the pole at $\{c_{2}-c_{1}\}$ {}{can exist} if $[c_{2}-c_{1}]$
is even {}{ and $c_2-c_1>2$}. In this case there are two subcases:
$[c_1]<-3$ and $[c_1]\geq -3$. If $[c_1]\geq-3$ it is a double
pole coinciding with the pole at $1-\{c_{2}+c_{1}\}$ if
$m_5(c_1)<c_2$, it is a simple pole if $-[c_1]-3<c_2\leq
m_5(c_1)$. If $[c_1]< -3$ it is a double pole coinciding with the
pole at $1-\{c_{2}+c_{1}\}$ if $-[c_1]-3<c_2$, and it is a simple
pole if $m_4(c_1)<c_2\leq -[c_1]-3$

\item the pole at $1-\{c_{2}-c_{1}\}$ {}{can exist} if $[c_{2}-c_{1}]$ is odd {}{ and $c_2-c_1<-2$},
it is a double pole coinciding with the pole at $\{c_{2}+c_{1}\}$
if $c_2<m_3(c_1)$, it is a simple pole if $m_3(c_1)\leq
c_2<-[c_1]$, and it is {} cancelled by {}{a} zero if $-[c_1]\leq
c_2$.

\item the pole at $\{c_{2}+c_{1}\}$ {}{can exist} if $[c_{2}+c_{1}]$ is even {}{ and $c_2+c_1<-1$},
it is a double pole coinciding with the pole at $1-\{c_2-c_1\}$ if
$c_2<m_3(c_1)$, it is a simple pole if $m_3(c_1)\leq c_2<[c_1]+2$,
and it is {} cancelled by {}{a} zero if $[c_1]+2\leq c_2$

\item the pole at $1-\{c_{2}+c_{1}\}$ {}{can exist} if $[c_{2}+c_{1}]$ is odd {}{ and $c_2+c_1>-1$},
it is a double pole coinciding with the pole at $\{c_2-c_1\}$ if
$m_5(c_1)<c_2$, it is a simple pole if $[c_1]-1<c_2\leq m_5(c_1)$,
and it is {} cancelled by {}{a} zero if $c_2\leq [c_1]-1$

\end{itemize}

For all the channels, {}{as soon as $\{c_2-c_1\}$ (respectively,
$\{c_2+c_1\}$) becomes integer, the poles at $\{c_2-c_1\}$ and
$1-\{c_2-c_1\}$ (respectively, $\{c_2+c_1\}$ and $1-\{c_2+c_1\}$)
fall out of the physical sheet. A separate analysis is however}
required for $c_1+c_2=0,-1,-2$, since the projectors $P_1,P_2,P_3$
are singular in this case.

{}{Let us notice that the presence of double poles in the physical
strip for specific values of the representation parameters might
be an indication of the Coleman-Thun mechanism
\cite{Coleman:1978kk}. One can also expect our R-matrix, which we
directly obtained form the Yangian construction, to be a bootstrap
R-matrix for particles in the fundamental three-dimensional
representation of $\mathfrak{sl}(2|1)$. The double poles we
observe should then be subject to a consistent multi-scattering
interpretation in the related bootstrap approach
\cite{Mussardo,Patrick}. We reserve this point for a further
investigation.}

\end{document}